\documentclass{iaus}
\usepackage{graphicx,epsf}

\def\la{\mathrel{\hbox{\rlap{\hbox{\lower4pt\hbox{$\sim$}}}\hbox{$<$}}}}

\title[A-Stars and the Virtual Observatory] 
{A-Stars and the Virtual Observatory}

\author[P. Padovani]   
{Paolo Padovani$^1$}

\affiliation{$^1$AVO Scientist, ST-ECF, European Southern Observatory,
             Karl-Schwarzschild-Str. 2, D-85748 Garching bei
             M\"unchen, Germany \break email: Paolo.Padovani@eso.org}

\pubyear{2004}
\volume{224}  
\pagerange{1--10}
\date{?? and in revised form ??}
\jname{The A-Star Puzzle}
\editors{J.\,Zverko, W.W.\,Weiss, J.\,\v{Z}i\v{z}\v{n}ovsk\'{y}, \& S.J.\,Adelman, eds.}
\begin{document}

\maketitle

\begin{abstract}
The Virtual Observatory (VO) will revolutionise the way we do Astronomy, by
allowing easy access to all astronomical data and by making the handling
and analysis of datasets at various locations across the globe much simpler
and faster. I report here on the need for the VO and its status in Europe,
including the first ever VO-based astronomical paper, and then give two
specific applications of VO tools to open problems of A-stars research.


\end{abstract}

\firstsection 

\section{Astronomy in the XXI century}

Astronomy is facing the need for radical changes. When dealing with surveys
of up to $\sim 1,000$ sources, one could apply for telescope time and
obtain an optical spectrum for each one of them to identify the whole
sample. Nowadays, we have to deal with huge surveys (e.g., the Sloan
Digital Sky Survey [SDSS; \cite{{Abazajian04}}], the Two Micron All Sky
Survey [2MASS; \cite{Cutri03}], the Massive Compact Halo Object [MACHO;
e.g., \cite{Alcock04}] survey), reaching (and surpassing) the 100 million
objects. Even at, say, 3,000 spectra at night, which is only feasible with
the most efficient multi-object spectrographs and for relatively bright
sources, such surveys would require more than 100 years to be completely
identified, a time which is clearly much longer than the life span of the
average astronomer!  But even taking a spectrum might not be enough to
classify an object. We are in fact reaching fainter and fainter sources,
routinely beyond the typical identification limits of the largest
telescopes available (approximately 25 magnitude for 2 - 4 hour exposures),
which makes ``classical'' identification problematic. These very large
surveys are also producing a huge amount of data: it would take about two
months to download at 1 Mbytes/s (an extremely good rate for most
astronomical institutions) the Data Release 2 (DR2; {\tt
http://www.sdss.org/dr2/}) SDSS images, about two weeks for the
catalogues. The images would fill up $\sim$ 1,000 DVDs ($\sim$ 500 if using
dual-layer technology). And the final SDSS will be about three times as
large as the DR2. These data, once downloaded, need also to be analysed,
which requires tools which may not be available locally and, given the
complexity of astronomical data, are different for different energy
ranges. Moreover, the breathtaking capabilities and ultra-high efficiency
of new ground- and space-based observatories have led to a ``data
explosion'', with astronomers world-wide accumulating $\approx 1$ Terabyte
of data per night. For example, the European Southern Observatory
(ESO)/Space Telescope European Coordinating Facility (ST-ECF) archive is
predicted to increase its size by two orders of magnitude in the next eight
years or so, reaching $\approx 1,000$ Terabytes. Finally, one would like to
be able to use all of these data, including multi-million-object
catalogues, by putting this huge amount of information together in a
coherent and relatively simple way, something which is impossible at
present.

All these hard, unescapable facts call for innovative solutions. For
example, the observing efficiency can be increased by a clever
pre-selection of the targets, which will require some ``data-mining'' to
characterise the sources' properties before hand, so that less time is
``wasted'' on sources which are not of the type under investigation. One
can expand this concept even further and provide a ``statistical''
identification of astronomical sources by using all the available,
multi-wavelength information without the need for a spectrum. The
data-download problem can be solved by doing the analysis where the data
reside. And finally, easy and clever access to all astronomical data
worldwide would certainly help in dealing with the data explosion and would
allow astronomers to take advantage of it in the best of ways.

\section{The Virtual Observatory}

The name of the solution is the Virtual Observatory (VO). The VO is an
innovative, evolving system, which will allow users to interrogate multiple
data centres in a seamless and transparent way, to utilise at best
astronomical data. Within the VO, data analysis tools and models,
appropriate to deal also with large data volumes, will be made more
accessible. New science will be enabled, by moving Astronomy beyond
``classical'' identification with the characterisation of the properties of
very faint sources by using all the available information. All this will
require good communication, that is the adoption of common standards and
protocols between data providers, tool users and developers. This is being
defined now using new international standards for data access and mining
protocols under the auspices of the recently formed International Virtual
Observatory Alliance (IVOA: {\tt http://ivoa.net}), a global collaboration
of the world's astronomical communities.

One could think that the VO will only be useful to astronomers who deal
with colossal surveys, huge teams and Terabytes of data! That is not the
case, for the following reason. The World Wide Web is equivalent to having
all the documents of the world inside one's computer, as they are all
reachable with a click of a mouse. Similarly, the VO will be like having
all the astronomical data of the world inside one's desktop. That will
clearly benefit not only professional astronomers but also anybody
interested in having a closer look at astronomical data. Consider the
following example: imagine one wants to find {\it all} high-resolution
spectra of A-type stars available in {\it all} astronomical archives in a
given wavelength range. One also needs to know which ones are in raw or
processed format, one wants to retrieve them and, if raw, one wants also to
have access to the tools to reduce them on-the-fly. At present, this is
extremely time consuming, if at all possible, and would require, even to
simply find out what is available, the use a variety of search interfaces,
all different from one another and located at different sites. The VO will
make it possible very easily.

\section{The VO in Europe: the Astrophysical Virtual Observatory}

The status of the VO in Europe is very good. In addition to seven current
national VO projects, the European funded collaborative Astrophysical
Virtual Observatory initiative (AVO: {\tt http://www.euro-vo.org}) is
creating the foundations of a regional scale infrastructure by conducting a
research and demonstration programme on the VO scientific requirements and
necessary technologies. The AVO has been jointly funded by the European
Commission (under the Fifth Framework Programme [FP5]) with six European
organisations participating in a three year Phase-A work programme. The
partner organisations are ESO in Munich, the European Space Agency,
AstroGrid (funded by PPARC as part of the United Kingdom's E-Science
programme), the CNRS-supported Centre de Donnees Astronomiques de
Strasbourg (CDS) and TERAPIX astronomical data centre at the Institut
d'Astrophysique in Paris, the University Louis Pasteur in Strasbourg, and
the Jodrell Bank Observatory of the Victoria University of Manchester. The
AVO is the definition and study phase leading towards the Euro-VO - the
development and deployment of a fully fledged operational VO for the
European astronomical research community. A Science Working Group was also
established two years ago to provide scientific advice to the project.

The AVO project is driven by its strategy of regular scientific
demonstrations of VO technology, held on an annual basis in coordination
with the IVOA. For this purpose progressively more complex AVO
demonstrators are being constructed. The current one, a downloadable Java
application, is an evolution of Aladin, developed at CDS, and has become a
set of various software components, provided by AVO and international
partners, which allows relatively easy access to remote data sets,
manipulation of image and catalogue data, and remote calculations in a
fashion similar to remote computing.

\section{Doing Science with the AVO}

The AVO held its second demonstration, 'AVO 1st Science', on January 27 -
28, 2004 at ESO. The demonstration was truly multi-wavelength, using
heterogeneous and complex data covering the whole electromagnetic
spectrum. These included: MERLIN, VLA (radio), ISO [spectra and images] and
2MASS (infrared), USNO, ESO 2.2m/WFI and VLT/FORS [spectra], and HST/ACS
(optical), XMM and Chandra (X-ray) data and catalogues. Two cases were
dealt with: an extragalactic case on obscured quasars, centred around the
Great Observatories Origin Deep Survey (GOODS) public data, and a Galactic
scenario on the classification of young stellar objects.

The extragalactic case was so successful that it turned into the first
published science result fully enabled via end-to-end use of VO tools and
systems, the discovery of $\sim 30$ high-power, supermassive black holes in
the centres of apparently normal looking galaxies (\cite[Padovani \etal\
2004]{Padovani04}). The AVO prototype made it much easier to classify the
sources we were interested in and to identify the previously known ones, as
we could easily integrate all available information from images, spectra,
and catalogues at once. This is proof that VO tools have evolved beyond the
demonstration level to become respectable research tools, as the VO is
already enabling astronomers to reach into new areas of parameter space
with relatively little effort. 

\section{The VO and A-type stars}

I have used the AVO prototype to tackle two problems of A-stars research,
namely, establishing membership to an open cluster and assessing if chemically
peculiar A-type stars are more likely to be X-ray emitters than normal A-type
stars. 

\subsection{Open cluster membership}

Cluster membership is vital to determine the distance, and therefore
absolute magnitude, and age of A-type stars, as discussed by Richard Monier
(\cite{Monierpoprad}) and Stefano Bagnulo (\cite{Bagnulopoprad}) at this
conference. In short, open clusters play a crucial role in stellar
astronomy because, as a consequence of the stars having a common age, they
provide excellent natural laboratories to test theoretical stellar
models. The AVO prototype can be of help in determining if a star does
belong to an open cluster, at various levels.

I have chosen the Pleiades as the target, since even extragalactic
astronomers like me know about it (although the value of its parallax is
strongly debated: see, e.g., \cite{Pan04})! I will also use this example to
describe the capabilities of the tool. A step-by-step guide which should
allow anyone to reproduce what I have done here with the AVO prototype can
be found at {\tt http://www.eso.org/$\sim$ppadovan/vo.html}.

\begin{figure}
\includegraphics[height=13cm]{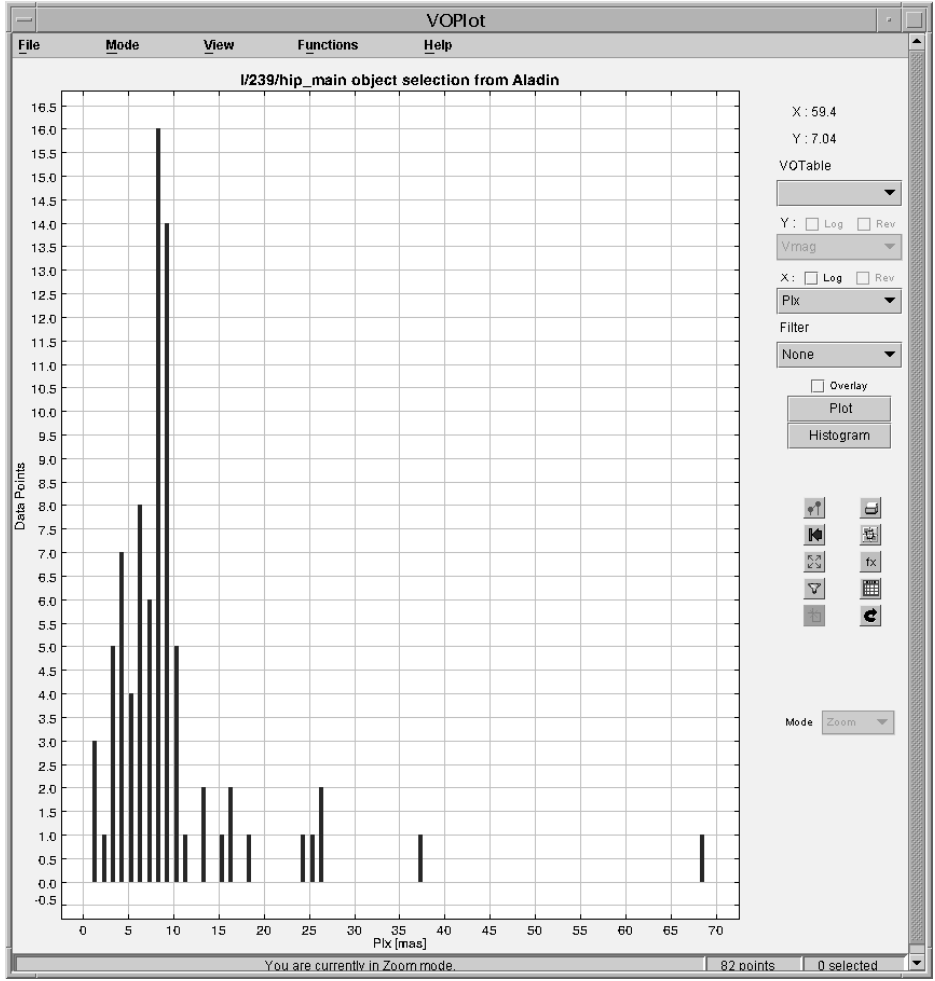} \caption{An histogram of the
Hipparcos parallaxes for the Pleiades done with VOPlot, the graphical
plug-in of the AVO prototype. Note the peak at around $8 - 9$ mas,
consistent with the cluster value of $8.46\pm0.22$ mas, and the presence of
many foreground and some background stars.}\label{fig:histo}
\end{figure}

\begin{figure}
\includegraphics[height=9.8cm]{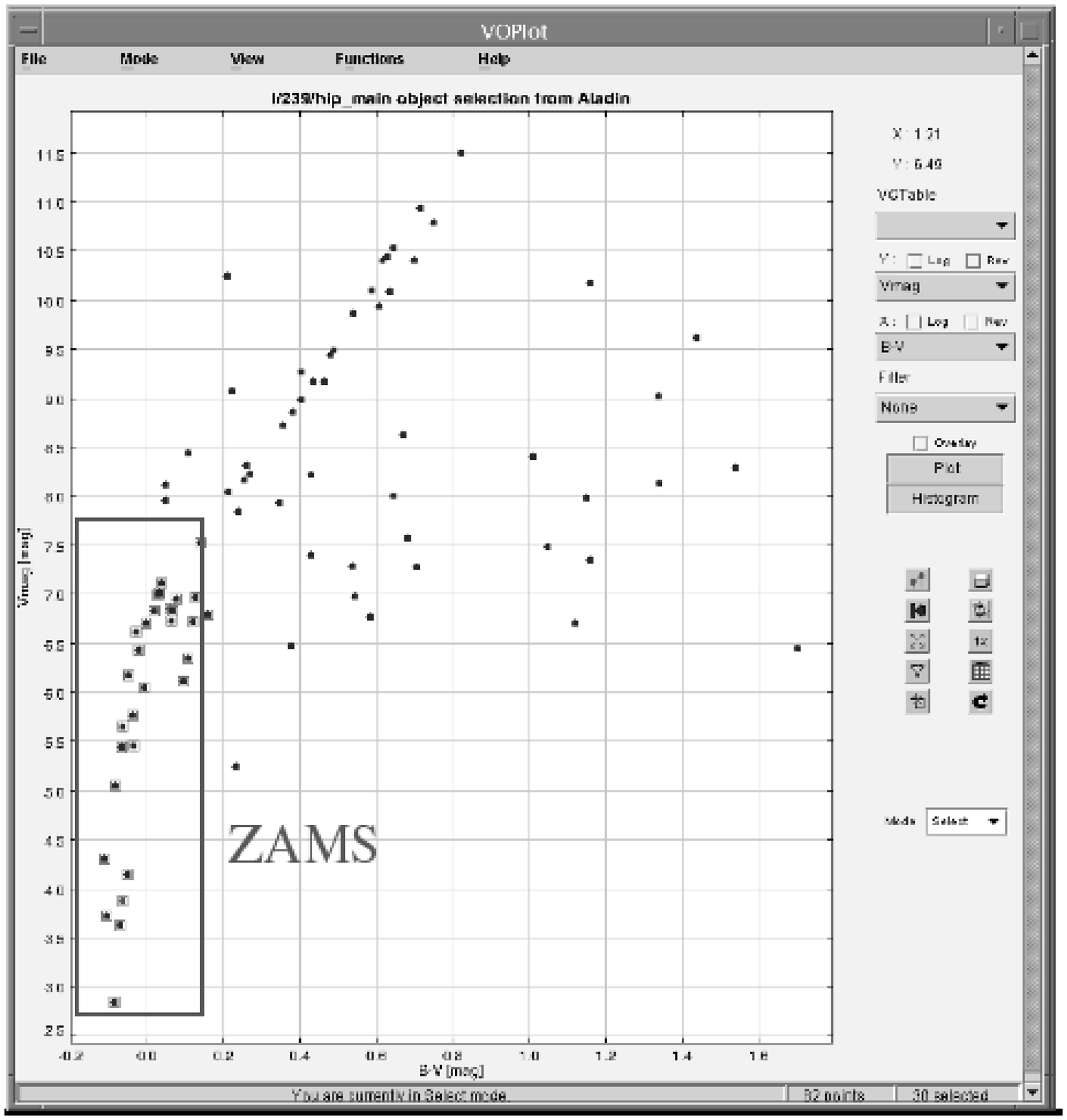}
\includegraphics[height=9.5cm]{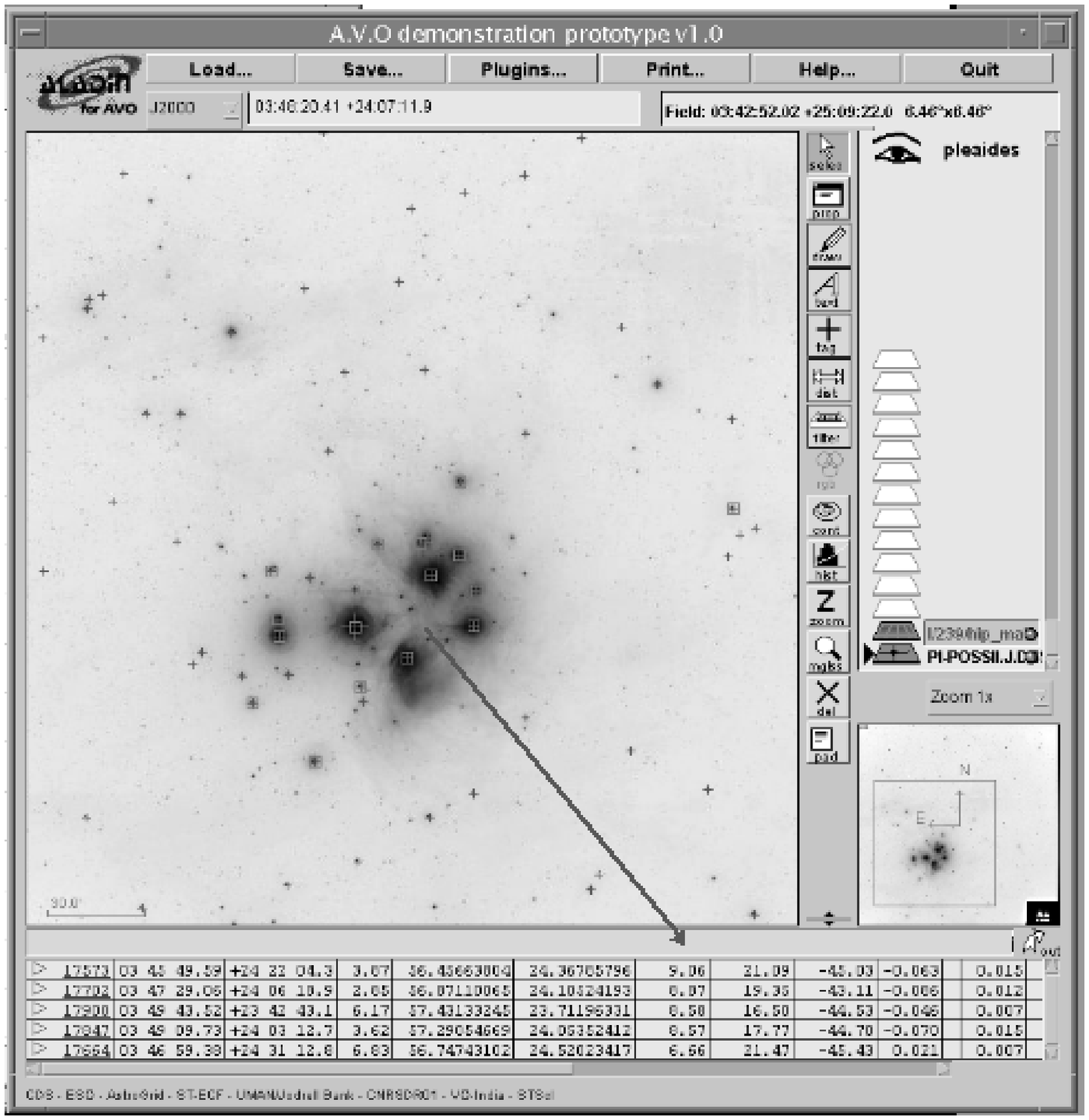}
\caption{Bright stars on the Zero Age Mean Sequence of the Pleiades
selected in VOPlot (top figure, bottom left corner) are highlighted
(squares) in the POSS II image (bottom figure) and turn out to be mostly in
the centre of the cluster, with parallaxes (indicated by the arrow) $\sim 8
- 9$ mas, consistent with the cluster value of $8.46\pm0.22$
mas.}\label{fig:zams}
\end{figure}

\begin{figure}
\includegraphics[height=9.7cm]{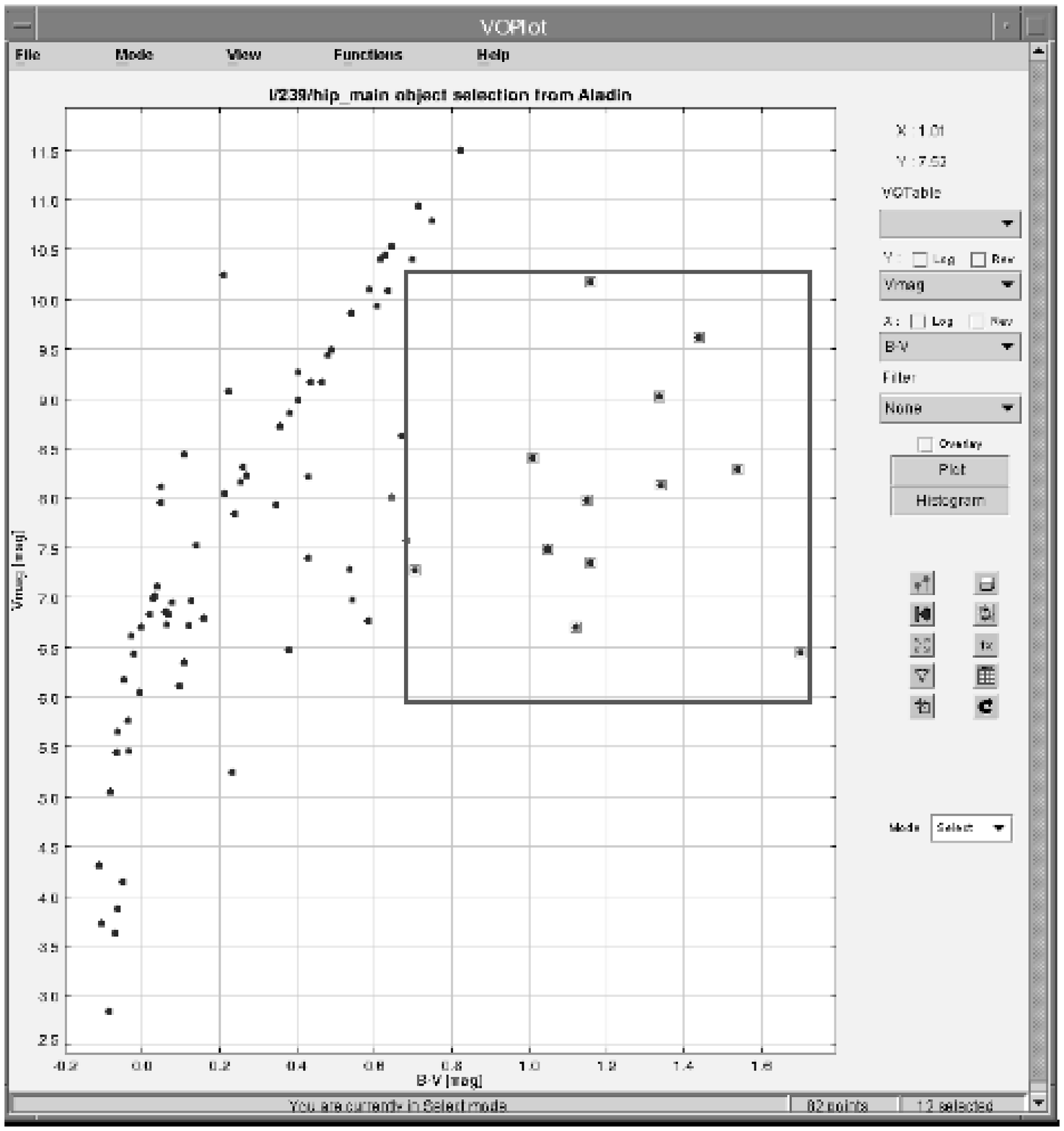}
\includegraphics[height=9.6cm]{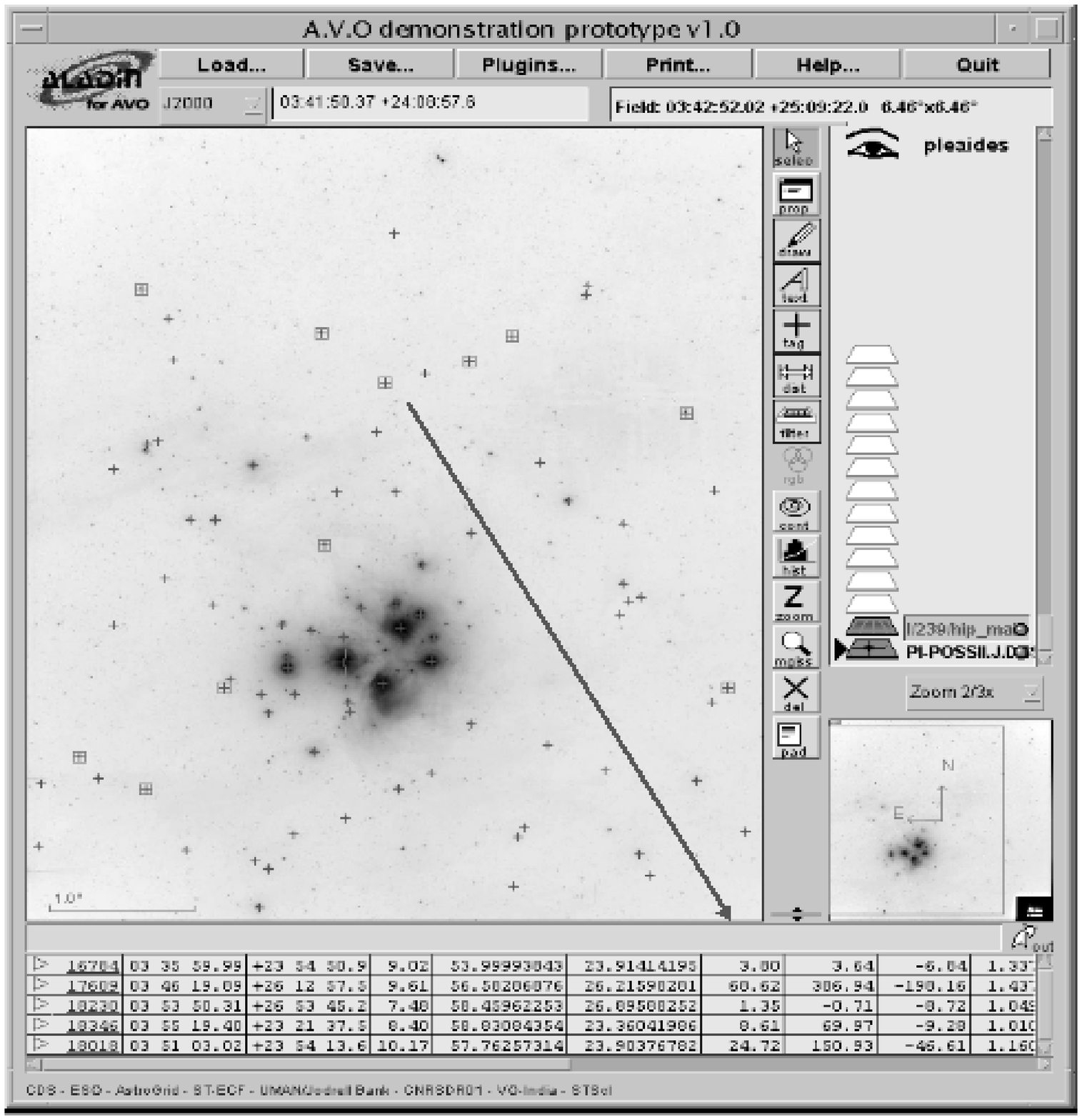}
\caption{Stars off the Zero Age Mean Sequence of the Pleiades selected in
VOPlot (top figure, top right corner) are highlighted (squares) in the POSS
II image (bottom figure) and turn out to be mostly at the outskirts of the
cluster, with parallaxes (indicated by the arrow) typically inconsistent
with the cluster value of $8.46\pm0.22$ mas.}\label{fig:nozams}
\end{figure}

We start by loading a Second Palomar Observatory Sky Survey (POSS II) image
of the Pleiades. Since we are interested in cluster membership, we need
distance information. The AVO prototype has a direct link to VizieR ({\tt
http://vizier.u-strasbg.fr/}), a service which provides access to the most
complete library of published astronomical catalogues and data tables
available on-line. We then search for all VizieR catalogues which provide
parallax information and find the Hipparcos catalogue (\cite{Perryman97}),
which we load into the prototype. All Hipparcos sources are automatically
overlaid on the POSS II image. We can now very easily plot an histogram of
the Hipparcos parallaxes by using VOPlot, the graphical plug-in of the
prototype. Most stars in the image have parallaxes $\sim 8 - 9$ mas,
consistent with the cluster value of $8.46\pm0.22$ mas (\cite[Robichon
\etal\ 1999]{Robichon99}), but there are also many foreground and some
background stars (Fig. \ref{fig:histo}). We now want to plot a colour --
magnitude diagram but first we need to correct for reddening (which is 0.04
for this cluster) the observed $(B-V)$ colour given in the Hipparcos
catalogue. We then create a new column, $(B-V)_{\rm o} = (B-V) - 0.04$, and
then plot the observed $V_{\rm mag}$ vs. $(B-V)_{\rm o}$. The Zero Age Mean
Sequence (ZAMS), flipped because we are using observed and not absolute
magnitudes, is clearly visible (Fig. \ref{fig:zams}, top). We can now have
a very nice, visual match between ZAMS and cluster membership. By selecting
in VOPlot bright stars which are on the ZAMS (top of Fig. \ref{fig:zams},
bottom left corner), the corresponding sources are highlighted in the
image. These are mostly in the centre of the cluster. By selecting them
with the cursor one can see that most of them have parallaxes $\sim 8 - 9$
mas, as expected (Fig. \ref{fig:zams}). On the other hand, if we now select
in VOPlot the sources off the ZAMS, in the POSS II image one can see that
they are mostly at the edge of the field and, looking at their parallaxes,
generally foreground sources, with a couple of background ones and only a
few possible members (Fig. \ref{fig:nozams}).

At this point one could do things properly, that is use the Hipparcos data
and the mean radial velocity of the cluster centre, together with eq. (1)
of \cite{Robichon99}, to add new columns with the relevant parameters to
the Hipparcos catalogue, and determine membership based on a statistical
criterion. At present, this would be quite cumbersome, although still
possible. Very soon, however, one will be able to add such functionality to
the VO as a ``Web Service''. Web Services will promote growth of the VO in
the way that web pages grew the World Wide Web. For example, a tool to
determine cluster membership could be ``published'' to the VO as a service
to which astronomers can send their input, in an appropriate format, and
then receive the output, e.g., a list of cluster members and non-members.
 
Using the AVO prototype, one can also look for data available for selected
sources at various mission archives. For example, in our case one could
select some sources in the image and then select ESO, the Hubble Space
Telescope (HST), and the International Ultraviolet Explorer (IUE) under
``Missions in VizieR''. The pointings for these three missions for the
sources under examination would then be overlaid in the image. Selecting
one of these pointings provides, in some cases (e.g., HST and IUE), links
to preview images, so that one can have a ``quick look'' at the archival
data. By clicking on the HST ``Dataset'' column the default Web browser
starts up and the dataset page at the Multimission Archive at STScI (MAST)
is made available. From there one can have access to all papers published
using those data. So from the AVO prototype one is only two ``clicks'' away
from the journal articles which have used the MAST data of the astronomical
sources in the image!

To get a flavour of the wealth of archival data available for A-type stars,
I have also cross-correlated the Sky2000 catalogue with the MAST holdings
using the service available at {\tt
http://archive.stsci.edu/search/sky2000.html}.  Out of $\sim 22,400$ A-type
stars, it turns out that 754 have IUE data, for a total of $\sim$ 10,000
observations, while 128 have non-IUE data (FOS, GHRS, STIS, FUSE, EUVE,
Copernicus, HUT, WUPPE, and BEFS; see the MAST site at {\tt
http://archive.stsci.edu/} for details on all of these missions), for a
total of 1,700 observations, $\sim 60\%$ of which are spectra. How many
more data are available in the other astronomical archives? Only the VO
will allow us to answer that question in a relatively simple way.

\subsection{Peculiar A stars in the X-ray band} 

The issue of the X-ray emission of A-type stars is a long standing one
(see, e.g, \cite[Simon, Drake, \& Kim 1995]{Simon95}). While it is clear
that these objects can be X-ray sources, it has been suggested that their
X-ray emission is not associated with the star itself but might come from a
binary companion. \cite{Damiani03} have analyzed {\it Chandra} observations
of the young open cluster NGC 2516 and detected only twelve A stars, out of
58, while six out of eight of the chemically peculiar (CP) A-stars were
detected (a difference significant at the $\sim 2\sigma$ level). It has
then been suggested, also on the basis of previous results (e.g.,
\cite{Dachs96}), that CP A-type stars are more easily detected in the
X-rays than normal A stars, although the astrophysical implications of this
result would not be straightforward. The AVO prototype provides a
relatively simple way to address directly and with sound statistics the
following question: ``Are CP A-type stars more likely to be X-ray emitters
than normal A-type stars?''.

\begin{figure}
\includegraphics[height=10cm]{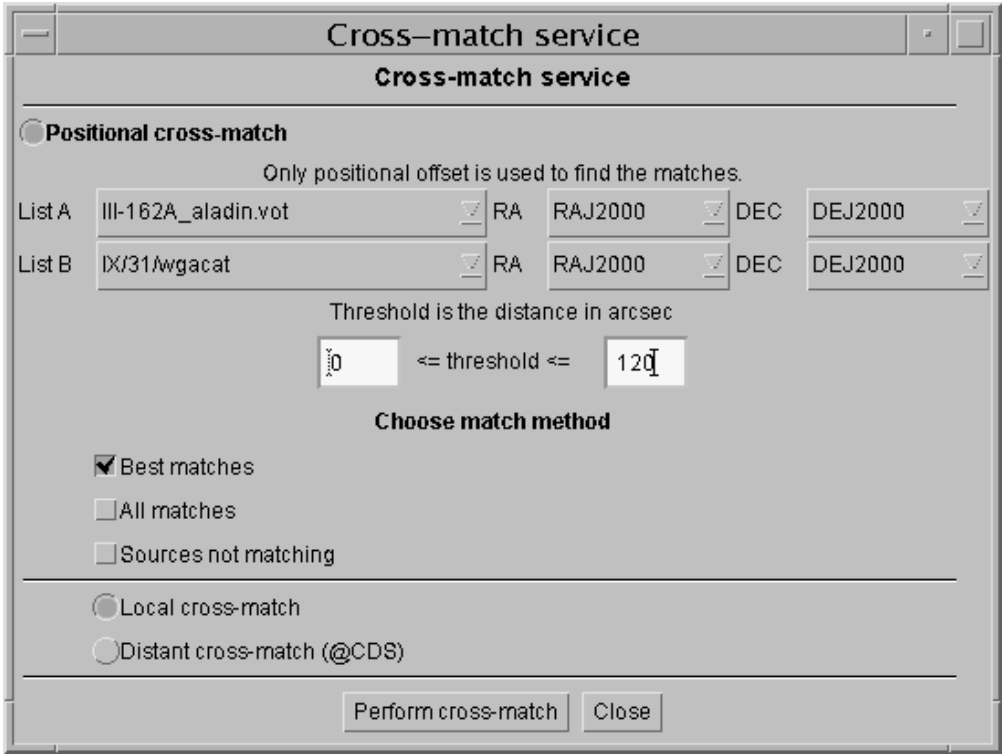} \caption{The cross-match
plug-in of the AVO prototype, used here to cross-correlate the
\cite{Renson91} catalogue of chemically peculiar stars with WGACAT, a
catalogue of all {\it ROSAT} observations.}\label{fig:cross}
\end{figure}

I first started by loading from VizieR the \cite{Renson91} catalogue of CP
stars, which include 6,684 sources, into the prototype. I then proceeded to
select the 4,736 A-type stars by using the ``Filter'' option. As comparison
samples I used the Henry Draper (HD) and SAO star catalogues, which contain
272,150 and 258,944 sources, out of which I selected the 72,154 and 47,230
A-stars respectively. I then loaded WGACAT (\cite[White, Giommi, \&
Angelini 1995]{whi95}), a catalogue of all {\it ROSAT} Position Sensitive
Proportional Counter (PSPC) observations, covering approximately 10\% of
the sky and including about 92,000 serendipitous sources (excluding the
targets, which could bias the result). The results of the
cross-correlations, done using the cross-match plug-in
(Fig. \ref{fig:cross}), are as follows (all errors are $3\sigma$ and
spurious sources have been subtracted off by doing the match with a shift
of 1 degree in the coordinates):

\begin{enumerate}

\item CP A-stars with WGACAT: 74/4736 matches, i.e., $1.6\pm0.6\%$; 

\item HD A-stars with WGACAT: 368/72154, i.e., $0.5\pm0.1\%$; correcting
statistically for CP stars contamination, assumed to be at the $10\%$ level
(e.g., \cite[Monin, Fabrika, \& Valyavin 2002]{Monin02}), one gets $\sim
0.4\%$;

\item SAO A-stars with WGACAT: 327/47230, i.e., $0.7\pm0.1\%$; correcting
statistically for CP stars contamination, one gets $\sim 0.6\%$.

\end{enumerate}

Therefore, one can conclude that {\it CP A-stars are 3 (SAO) to 4 (HD)
times more likely to be X-ray sources than normal A-type stars} with very
high significance. Selecting only the magnetic stars in \cite{Renson91},
which can be identified has having Sr, Cr, Eu, Si, He, Ti, and Ca
classification peculiarities (e.g., \cite{Landstreet92}), one finds 22/1433
matches, that is a $1.5\pm1.0\%$ detection rate, not significantly
different from that of all CP A-stars. It then appears that the presence of
a magnetic field does not play a role in triggering X-ray emission, a
somewhat puzzling result which deserves further investigation.

A related issue is that of radio emission. The strong fields present in the
magnetic subclass of CP stars (see above), in fact, should give rise to
radio emission, for example via the gyrosynchrotron mechanism
(\cite{Trigilio04}). I have then cross-correlated the A-type stars in the
\cite{Renson91} catalogue with two large-area radio catalogues, namely the
NRAO-VLA Sky Survey (NVSS; \cite{Condon98}), which covers the sky north of
$\delta = -40^{\circ}$ down to $\sim 3.5$ mJy at 1.4 GHz, and the Faint
Images of the Radio Sky at Twenty-centimeters (FIRST; \cite{White97}),
which covers $\sim 1/5$ of the sky, mostly in the north Galactic cap, down
to $\sim 1$ mJy 1.4 GHz. I found no matches. In retrospect, this is not
surprising as the few CP stars detected so far have radio fluxes $\la 1 -
2$ mJy (e.g., \cite{Drake87}).

\section{Conclusions}\label{sec:concl}

The main conclusions are as follows:

\begin{enumerate}

\item We need to change the way we do Astronomy if we want to take
advantage of the huge amount of data we are being flooded with. The way to
do that is through the Virtual Observatory.

\item The Virtual Observatory will make the handling and analysis of
astronomical data and tools located around the world much easier, enabling
also new science.

\item Everybody will benefit, including A-type star researchers!

\item Virtual Observatory tools are available now to facilitate
astronomical research and, as I have shown, can also be applied to A-stars.

\end{enumerate}

Visit {\tt http://www.euro-vo.org/twiki/\-bin/\-view/Avo/SwgDownload}
to download the AVO prototype. I encourage astronomers to download the
prototype, test it, and use it for their own research. For any
problems with the installation and any requests, questions, feedback,
and comments you might have please contact the AVO team at
twiki@euro-vo.org. (Please note that this is still a prototype:
although some components are pretty robust some others are not.)

\begin{acknowledgments}
I would like to thank the organizers of the conference for their kind
invitation, which has allowed an extragalactic astronomer like me to learn
about stars! I am also grateful to Stefano Bagnulo for his help in
preparing my talk and to Mark Allen for reading this paper. I have made
extensive use of the CDS VizieR catalogue tool, SIMBAD and the Aladin sky
atlas service. The Astrophysical Virtual Observatory was selected for
funding by the Fifth Framework Programme of the European Community for
research, technological development and demonstration activities, under
contract HPRI-CT-2001-50030.
\end{acknowledgments}

\begin{discussion}

\discuss{Skoda}{How will you manage copyright rules, that is, who should one
cite in research done using the VO?}

\discuss{Padovani}{The original data providers, as far as the data are
concerned, and any relevant VO project for the tools.}

\discuss{Skoda}{What if I ask for some data (e.g., spectra of a particular
star) which are still under proprietary period? Will I be informed about
this?}

\discuss{Padovani}{Yes. You will be told that the data are present but
still proprietary. You will also be told when you can access them and
perhaps even reminded when they become available. But remember that the VO
is still not fully operational so at present you simply cannot access the
data.}

\discuss{Kub\'at}{How do you control the quality of the data in catalogues?}

\discuss{Padovani}{That is a very important point but the VO cannot do
that, as it would require a lot of resources. Data quality control will be
up to those who know the data best, that is the data providers.} 

\end{discussion}


\begin{thebibliography}{}

\bibitem[Abazajian \etal\ 2004]{Abazajian04}
     {Abazajian, K. et al.} 2004, 
     \textit{AJ} 128, 502 

\bibitem[Alcock \etal\ 2001]{Alcock04}
     {Alcock, C. et al.} 2001,  
     \textit{ApJS} 136, 439 

\bibitem[2004]{Bagnulopoprad}
     {Bagnulo, S.} 2004, 
     \textit{These proceedings}


\bibitem[Condon \etal\ 1998]{Condon98}
     {Condon, J.J., Cotton, W.D., Greisen, E.W., Yin, Q.F., Perley, R.A., 
     Taylor, G.B. \& Broderick, J.J.} 1998, 
     \textit{AJ} 115, 169

\bibitem[Cutri \etal\ 2003]{Cutri03} 
     {Cutri, R. M. et al.} 2003, 
     \textit{Explanatory Supplement to the 2MASS Second Incremental Data
     Release} available at 
     {\tt http://www.ipac.caltech.edu/\-2mass/\-releases/\-allsky/\-doc/\-explsup.html}

\bibitem[Dachs \& Hummel 1996]{Dachs96}
     {Dachs, J. \& Hummel, W.} 1996,
     \textit{A\&A} 312, 818

\bibitem[Damiani \etal\ (2003)]{Damiani03} 
     {Damiani, F., Flaccomio, E., Micela, G., Sciortino, S., Harnden,
     F.R., Murray, S.S., Wolk, S.J. \& Jeffries, R.D.} 2003, 
     \textit{ApJ} 588, 1009

\bibitem[Drake \etal\ 1987]{Drake87}
     {Drake, S.A., Abbott, D.C., Bastian, T.S., Bieging, J.H., 
      Churchwell, E., Dulk, G. \& Linsky, J.L.} 1987, 
     \textit{ApJ} 322, 902 

\bibitem[Landstreet 1992]{Landstreet92} 
     {Landstreet, J.D.} 1992, 
     \textit{A\&AR} 4, 35 

\bibitem[2004]{Monierpoprad}
     {Monier, R. \& Richard, O.} 2004,
     \textit{These proceedings}

\bibitem[Monin, Fabrika, \& Valyavin (2002)]{Monin02}
     {Monin, D.N., Fabrika, S.N. \& Valyavin, G.G.} 2002,
     \textit{A\&A} 396, 131

\bibitem[Padovani \etal\ (2004)]{Padovani04}
     {Padovani, P., Allen, M.G., Rosati, P. \& Walton, N.A.} 2004,
     \textit{A\&A} in press (astro-ph/0406056) 

\bibitem[Pan, Shao, \& Kulkarni 2004]{Pan04}
     {Pan, X., Shao, M., \& Kulkarni, S.R.} 2004, 
     \textit{Nature} 427, 326

\bibitem[Perryman 1997]{Perryman97}
     {Perryman, M. A. C.} 1997,
     \textit{The Hipparcos and Tycho Catalogues}, ESA SP-1200 
     (Noordwijk: ESA) 

\bibitem[Renson, Gerbaldi \& Catalano (1991)]{Renson91}
     {Renson, P., Gerbaldi, M. \& Catalano, F.A.} 1991,
     \textit{A\&AS} 89, 429 

\bibitem[Robichon \etal\ (1999)]{Robichon99}
     {Robichon, N., Arenou, F., Mermilliod, J.-C. \& Turon, C.} 1999,
     \textit{A\&A} 345, 471 

\bibitem[Simon, Drake, \& Kim (1995)]{Simon95}
     {Simon, T., Drake, S.A. \& Kim, P.D.} 1995,
     \textit{PASP} 107, 1034

\bibitem[Trigilio \etal\ 2004]{Trigilio04}
     {Trigilio, C., Leto, P., Umana, G., Leone, F. \& Buemi, C.S.} 2004,
     \textit{A\&A} 418, 593 
     
\bibitem[White, Giommi, \& Angelini (1995)]{whi95}
     {White, N.E., Giommi, P. \& Angelini, L.} 1995, 
     on-line database at {\tt http://wgacat.gsfc.nasa.gov}

\bibitem[White \etal\ 1997]{White97}
     {White, R.L., Becker, R.H., Helfand, D.J. \& Gregg, M.D.} 1997, 
     \textit{ApJ} 475, 479 

\end{thebibliography}
\end{document}